# Controlling Interface States in 1D Photonic Crystals by tuning Bulk Geometric Phases

Wensheng Gao,[1] Meng Xiao,[2] Baojie Chen,[3] Edwin Y. B. Pun,[3] C. T. Chan,[1,2] and Wing Yim Tam[1,*]

[1)]*Department of Physics and William Mong Institute of Nano Science and Technology*
*Hong Kong University of Science and Technology*
*Clear Water Bay, Kowloon, Hong Kong, China*

[2)]*Department of Physics and Institute for Advanced Study, Hong Kong University of Science and Technology, Clear Water Bay, Kowloon, Hong Kong, China*

[3)]*Department of Electric Engineering, City University of Hong Kong, Hong Kong, China*

## Abstract

Interface states in photonic crystals usually require defects or surface/interface decorations. We show here that one can control interface states in 1D photonic crystals through the engineering of geometrical phase such that interface states can be guaranteed in even or odd, or in all photonic bandgaps. We verify experimentally the designed interface states in 1D multilayered photonic crystals fabricated by electron beam vapor deposition. We also obtain the geometrical phases by measuring the reflection phases at the bandgaps of the PCs and achieve good agreement with the theory. Our approach could provide a platform for the design of using interface states in photonic crystals for nonlinear optic, sensing, and lasing applications

* Corresponding Author: phtam@ust.hk; Phone: 852-2358-7490; Fax: 852-2358-1652

**1. Introduction**

Interface states serve an important role in modern science and technology, from semiconductors [1, 2] and, recently, to photonic crystals [3]. Interface states in photonic crystals have been intensively applied to enhance nonlinear optical effects [4-6] and interface state lasing [7, 8], and also to related areas such as optical trapping [9], giant Goos-hänchen effect [10] and optical sensing [11]. Such effects are dramatically enhanced due to field localizations and order of magnitude enhancements of field amplitudes at the interface. A common technique to generate interface states is by adding defect layers or decorations to the photonic crystals. Interface states can then be achieved by tuning the parameters (materials or dimensions) of the defects/decorations. However, fine tuning is required as the interface states are sensitive to the details of the defects such that small variations may destroy the interface states [12].

It is known that the existence of interface state is related to the surface impedance. When two semi-infinite photonic crystals meet at an interface, one on the left hand side (LHS) and the other on the right hand side (RHS) of the interface, the condition for the existence of interface state requires the sum of the surface impedance to be zero [13, 14], i.e.

$$Z_L + Z_R = 0, \qquad (1)$$

where $Z_L$ and $Z_R$ are the impedance of the LHS and RHS photonic crystals, respectively. That is the surface impedances should have equal amplitudes but opposite in signs. Equation (1) provides a simple guide line to obtain interface states. Furthermore, it is known that the surface property (impedance/reflection phase) of a photonic crystal can be related to its bulk property (geometric/Zak phase) such that interface states can be guaranteed by proper design for the photonic bands of the crystal [14]. Based on this approach, we have designed 1D photonic crystals with interface states selectively in only even or odd, or in all, bandgaps of the photonic crystals. To demonstrate our approach, we fabricate photonic crystals with the designed interface states for the optical range. In addition, we have also measured the reflection phase, and hence the Zak phase, of the photonic crystals using a thick-gap Fabry-Perot interferometry technique [15, 16] as reported recently in the measurements of the Zak phase of 1D photonic crystals [17]. Our approach could potentially facilitate the design of photonic crystals in applications like nonlinear optics, sensors, and lasers.

## 2. Design of photonic crystals with interface state

*2.1 Reflection phase and Zak phase*

The surface impedance in Eq. (1) can be expressed in terms of the reflection coefficients $r_L$ and $r_R$ of the LHS and RHS photonic crystals as:

$$Z_L = \frac{1+r_L}{1-r_L} Z_0 \text{ and } Z_R = \frac{1+r_R}{1-r_R} Z_0, \qquad (2)$$

where $Z_0 = \sqrt{\frac{\mu_0}{\varepsilon_0}}$ is the vacuum impedance. For materials without losses the reflection coefficient can be expressed as $r_L = e^{i\phi_L}$, where $\phi_L$ is the reflection phase of the LHS photonic crystal. (Similarly, $r_R = e^{i\phi_R}$ for the RHS photonic crystal.) Thus the condition for interface state as given in Eq. (1) can be rewritten as $r_L r_R = 1$, or $\phi_L + \phi_R = 2m\pi$ for natural number *m*. By a proper choice for the reflection phase *m* can take the zero value, leading to a simpler condition for the reflection phase, i.e. $\phi_L = -\phi_R$. Note that $Z_L/Z_0$ (similarly for $Z_R/Z_0$) is a pure imaginary number inside a bandgap, i.e. $Z_L/Z_0 = (1+r_L)/(1-r_L) = i\zeta_L$, for real number $\zeta_L$. Thus the reflection phase can be expressed as $\phi_L = \pi - 2\arctan(\zeta_L)$, leading to $\text{sgn}[\phi_L] = \text{sgn}[\zeta_L]$ and, similarly, $\text{sgn}[\phi_R] = \text{sgn}[\zeta_R]$. (Here we chose the reflection phase inside the range $[-\pi, \pi)$.)

It is shown that for a photonic crystal consisting of two components (slabs A and B having thickness $d_a$ and $d_b$, respectively, for a total unit cell thickness $\Lambda = d_a + d_b$) with inversion symmetry, the reflection phase is related to the bulk geometric phase (Zak phase $\theta^{Zak}$ for 1D system) [14]. Furthermore, the reflection phases at the bandgaps of such a photonic crystal vary continuously either from $-\pi$ to 0 or from 0 to $\pi$. Hence only the signs of the reflection phases are needed for the determination of the existence of interface states. For bands and bandgaps below the first band crossing of the photonic crystal, the expression for the phase relation can be written as [14]:

$$\text{sgn}(\phi_n) = \text{sgn}(\zeta_n) = (-1)^n \exp(i\sum_{m=0}^{n-1}\theta_m^{Zak}), \qquad (3)$$

for all bands and bandgaps below the $n^{th}$ bandgap, and $\theta_m^{Zak} = 0$ or $\pi$. Here the Zak phase of the lowest $0^{th}$ band is expressed as:

$$\exp(i\theta_0^{Zak}) = \text{sgn}[1 - \frac{\epsilon_a \mu_b}{\epsilon_b \mu_a}], \qquad (4)$$

where $\varepsilon_a$, $\mu_a$ and $\varepsilon_b$, $\mu_b$ are the relative permittivity and permeability of the slabs A and B, respectively. Thus, for nonmagnetic dielectric materials $\mu_a = \mu_b = 1$, the sign of $\theta_0^{Zak}$ is determined by the permittivities of the slabs. For photonic crystal with an inversion center at the low refractive index slab (Type I), i.e. $\varepsilon_a < \varepsilon_b$, $\theta_0^{Zak}$ is 0; otherwise ($\varepsilon_a > \varepsilon_b$) it is π for photonic crystal with inversion center at the high refractive index slab (Type I'). Hence the center of the inversion is an important parameter for tuning interface states, and by designing the Zak phases at higher bands, one can determine the reflection phases of the bandgaps and hence the interface state at each bandgap.

The Zak phase at higher bands can be determined by the condition:

$$\sin\left(\frac{2\pi n_i d_i}{\lambda}\right) = 0, \tag{5}$$

where i = a or b for inversion center taken in slab B or A, respectively [14]. Here $n_i = \sqrt{\epsilon_i \mu_i}$ is the corresponding refractive index for slab A or B, and $\lambda$ is the wavelength. The Zak phase of an isolated band will be $\pi$ if $\lambda$ falls inside the band; otherwise, it is 0 [14].

*2.2 Photonic crystals with interface states*

Table I shows the properties of three composite photonic crystals (PCs) designed with photonic bands and interface states for the optical range. We chose $SiO_2$ and $TiO_2$ for the first two PCs, PC_I and PC_II, because these materials have relatively large refractive index contrast and, more importantly, negligible absorptions in the frequency range of interest (wave number 0.4 µm$^{-1}$ to 1.2 µm$^{-1}$). For PC_III, we use $SiO_2$ and Si as the material components for the IR range.

The first composite PC, PC_I (PC_A'A), consisting of photonic crystal PC_A' on photonic crystal PC_A, is a straight forward design by using a single PC with the same material components A and B ($SiO_2$ and $TiO_2$ respectively) and same dimensions ($d_a$ and $d_b$), but different $0^{th}$ band Zak phase due to inversion center with Type I or Type I' symmetry. PC_A has inversion center at the low refractive index material, here $SiO_2$, while PC_A' has inversion center at the high refractive index material, here $TiO_2$. This design guarantees the overlap of photonic bands and bandgaps. The Zak phases of the first 4 bands for PC_A, as determined below, are listed in the second column and the reflection phases of the bandgaps obtained from Eq. (3) are shown in the third column of Table I. By default, the Zak phases of bands above the $0^{th}$ band for PC_A' are opposite to those of PC_A as shown in the $4^{th}$ column of Table I. Since now the Zak phase of the $0^{th}$ band of PC_A' is also opposite to that of PC_A, the reflection phases for PC_A' listed in column 5 of Table I can differ from those of PC_A. Thus PC_I (PC_A'A) will exhibit interface state at odd bandgaps as the reflection phases are opposite for the corresponding bandgaps as shown in $6^{th}$ column of Table I.

The second composite PC, PC_II (PC_BA), consists of the same PC_A as in PC_I but PC_A' is now replaced by PC_B which has the same inversion center as PC_A. In order to have overlapping bands and bandgaps, the thicknesses of the components, $SiO_2$ and $TiO_2$, for PC_B are tuned by keeping the optical path $n_a d_a + n_b d_b$ the same as that of PC_A. Furthermore, the Zak phases of bands above the $0^{th}$ band are chosen the same as those of PC_A', as shown in the $9^{th}$ column of Table I. Now due to the fact that the Zak phase of the $0^{th}$ band of PC_B is the same as that of PC_A, the signs of the reflection phases for the bandgaps of PC_B, column $10^{th}$, are now opposite to those of PC_A'. This leads to interface states of the composite PC_BA occurring only at the even bandgaps as shown in the $11^{th}$ column in Table I.

The third composite PC, PC_III (PC_D'C), is designed such that the reflection phases are all opposite for all bandgaps of PC_C and PC_D' below the first band crossing. To achieve this we replace $TiO_2$ with Si. However, now the band crossing occurs after the second band, and thus only the first two bandgaps in the IR range can be studied. The Zak phases for PC_C and PC_D' are shown in columns $12^{th}$ and $14^{th}$ of Table I, respectively. The reflection phases for the bandgaps of the two PCs are shown in columns $13^{th}$ and $15^{th}$ of Table I. Since the signs of the reflection phases for the corresponding bandgaps are all opposite, interface states are thus guaranteed in all bandgaps.

Figure 1 shows the band diagrams (wavenumber $1/\lambda$ vs. $q\Lambda/2\pi$, $q$ is the wave vector) obtained analytically using parameters as stated in Table I for all the PCs as discussed in above. The Zak phases of the bands, obtained using Eqs. [4] and [5], are labeled in red. The horizontal red dash lines in Fig. 1 correspond to bands with Zak phase equal to $\pi$ as determined by Eq. [5]. For bands without the red dash line, the Zak phases are zero. We also color code each band to show the sign of the reflection phase: cyan for positive and purple for negative reflection phase, respectively. Now, it is easy to identify the bandgaps of the composite PCs that exhibit interface states, simply by requiring different colors of the corresponding bandgaps.

Note that the position of the interface state at a bandgap of the composite PC depends mainly on the dispersions of the materials while the interface state depends only on the geometrical/topological properties for the bulk bands of PC which is robust against perturbations. Thus by careful design, one can have interface states at harmonics of the incident light, e.g. second or third, such that they can be applied to nonlinear optics or sensors.

3**. Experiment**

*3.1 Sample fabrication and characterization*

In our experiment, one unit cell of Type I photonic crystals with the inversion center chosen at the low refractive index material, i.e. PC_A, PC_B, and PC_C, contains two half layers of SiO$_2$ (thickness $d_a/2$) sandwiching one layer of TiO$_2$ or Si (thickness $d_b$). As for the Type I' PCs with inversion center chosen at the high refractive index material, i.e. PC_A' and PC_D', one unit cell contains two half layers of TiO$_2$ or Si (thickness $d_a/2$) sandwiching one layer of SiO$_2$ (thickness $d_b$). We used a Cooke e-beam evaporation system (vacuum ~$10^{-6}$ mbar, deposition rates 0.15 nm/s and 0.3 nm/s for TiO$_2$ and SiO$_2$, respectively) to deposit the SiO$_2$ and TiO$_2$ layers for PC_A, PC_A' and PC_B. For PC_C and PC_D' we used another e-beam system (JunSun EBS-500 Dual Source Thermal / Electron Beam Evaporator, at ~0.5nm/s rate) and a thermal evaporator (Denton vacuum DV-502A, at also ~0.5nm/s rate) for the $SiO_2$ and Si layers, respectively. We first deposited a PC on a glass substrate and subsequently, with half of the first PC covered by another glass plate, we deposited a second PC on the top of first PC as well as on the covering glass to produce three PCs: one composite PC and two individual PCs. Finally we deposited a strip of ~200nm thick Ti by a Peva 450 EL e-beam evaporation system onto the individual PCs as well as the composite PC for reflection phase measurements. Note that, the photonic crystal PC_II (PC_BA) has been reported recently for the measurements of Zak phase [17].

We fabricated 4 to 6 unit cells photonic crystals using the dimensions as provided in Table I. Figure 2 shows cross section Scanning Electron Microscope (SEM) images of the PCs fabricated. The materials for the PCs are as labeled and the average thicknesses are given in the figure caption. Note that the resulted dimensions of the PCs are slightly different from those given in Table I. Furthermore, PC_III has slightly larger thickness variations as compared to PC_I and PC_II. Despite the small number of layers, we are able to obtain over-lapping bandgaps (first two rows) and, more importantly, the interface states (last row), shown as green curves in Fig. 3, the reflectance for all the PCs using a 5X objective microscopic system [16]. The bandgaps for the PCs are color coded for easy identification. The 4 vertical black lines, one for PC_I and PC_II, and two for PC_III, in Fig. 3 mark the locations of the interface states. The observed interface states are in good accord with the predictions shown in Table 1.

*3.2 Determination of reflection phase*

We followed the same procedures as reported earlier, using a thick-gap Fabry-Perot technique, to determine the reflection phases, and hence the Zak phases, of the PCs [16, 17]. To measure the reflection phase we placed a thick (~1cm) optically flat glass at ~10-20μm above the PC to form an air-gap Fabry-Perot etalon. We recorded the peak and trough wavelengths for the reflectance of the PC etalon, shown as blue curves in the first two rows of Fig. 3, and from which obtained the reflection phase using the constructive and destructive interference conditions as descripted in Ref. [16]. We then repeated the same procedures for the Ti layer coated on the same PC sample in the same setting to obtain the reflection phase of Ti. From the two measurements, we then obtained the relative reflection phase shift of the PC with respect to that of Ti to eliminate the numerical aperture effect of the objective [19]. Finally, using the known reflection phase of Ti [18] we obtained the reflection phase of the PC. The results of the reflection phases are shown as black circles in the first two rows of Fig. 3. Note that the signs of the reflection phases of the corresponding PCs are opposite at the interface state(s) as guided by the vertical black lines in the $1^{st}$, $4^{th}$, $5^{th}$ and $6^{th}$ columns of Fig. 3. Furthermore, the absolute reflection phases at the interface states are roughly equal in accord with the prediction for the reflection phase stated in above.

To further verify our experimental results, we have also performed calculations for the reflectance for the finite sized PCs, 4-6 unit cells, using the Transfer Matrix method by neglecting the losses from the materials of the PCs. We used the experimental parameters, with some adjustments to account for the variations in the thicknesses as obtained from the SEM images in Fig. 2 and the difference in the locations for the reflection phase and SEM measurements, in our calculations. The results for the calculated reflectance for the composite PCs are shown as black curves in the last rows of Fig. 3. The overall agreement of the calculated bandgaps and also the interface states for PCs with the experiment is good, despite that there are shifts of the interface states for PC_III (see the last two columns of Fig. 3) due possibly to the larger variations in the layer thickness of the PC. More importantly, the calculated reflection phases for the individual PCs are consistent (black curves in the first two rows of Fig.3) with the experiment.

## 4. Conclusion

Based on the intimate relation between the surface impedance and bulk geometrical phase, we design photonic crystals that exhibit interface states at designated bandgaps for the optical range. We verify our approach by fabricating the photonic crystals with the designed interface states. Furthermore, we have also measured the reflection phases, and hence the geometrical Zak phases, using a thick gap Fabry-Perot technique and the results are in good agreement with the calculations. Our approach could potentially lead to applications of interface state in photonic crystals for nonlinear optics or related areas.

**Acknowledgments**

Support from Hong Kong RGC grants AoE P-02/12 is gratefully acknowledged. The technical support of the Raith-HKUST Nanotechnology Laboratory for the electron-beam lithography facility at MCPF (SEG_HKUST08) is hereby acknowledged.

**Reference:**

[1]  A. Goetzberger, E. Klausmann, and M. J. Schulz, "Interface states on semiconductor/insulator surfaces," *CRC Critical Reviews in Solid State Sciences*, **6**, 1-43, 1976.

[2]  A. G. Milnes, "Semiconductor Devices and Integrated Electronics," Springer, Van Nostrand Reinhold Co. New York, 1980.

[3]  J. D. Joannopoulos, S. G. Johnson, J. N. Winn, and R. D. Meade "Photonic Crystals: Molding the Flow of Light," Princeton University Press, 2008.

[4]  J. Y. Ye, M. Ishikawa, Y. Yamane, N. Tsurumachi, and H. Nakatsuka, "Enhancement of two-photon excited fluorescene using one-dimensional photonic crystals," *Appl. Phy. Lett.* **75**, 3605, (1999).

[5]  H. Inouye and Y. Kanemitsu, "Direct observation of nonlinear effects in a one-dimensional photonic crystal," *Appl. Phy. Lett.* **82**,1155, (2003).

[6]  G. Ma and S. H. Tang, J. Shen, Z. Zhang, and Z. Hua, "Defect-mode dependence of two-photon-absorption enhancement in a one-dimensional photonic bandgap structure," *Opt. Lett.* **29**,1769, (2004).

[7]  L. Florescu, K. Busch, and S. John, "Semiclassical theory of lasing in photonic crystals," *J. Opt. Soc. B19*, 2215-2223 (2002).

[8]  M. H.Kok, W. Lu, J. C. W. Lee, W. Y. Tam, G. K. L. Wong and C. T. Chan, "Lasing From Dye-doped Photonic Crystals With Graded Layers in Dichromate Gelatin Emulsions," *Appl. Phy. Lett.* **92**, 151108/1-3 (2008).

[9]  D. A.Shilkin, E. V. Lyubin, I. V. Soboleve, and A. A. Fedyanin, "Direct measurements of forces induced by Bloch surface waves in a one –dimensional photonic crystals," *Optics Lett.* 40, 4883, (2015).

[10] I. V. Soboleva, V. V. Moskalenko, and A. A. Fedyanin, "Giant Goos-Hänchen effect and Fano resonance at photonic crystal surfaces," *Phy. Rev. Lett.,* **108**, 123901 (2012).

[11] F. Villa and L. E. Regalado, F. R.Mendieta and J. Gaspar-Armenta, T. Lopez-Rios, "Photonic crystal sensor based on surface waves for thin-film characterization," *Opt.Lett.* **27**, 646 (2002).

[12] S. Feng, H. Y. Sang, Z. Y. Li, B. Y. Cheng and D. Z. Zhang, "Sensitivity of surface states to the stack sequence of one-dimensional photonic crystals," *J. Opt. A: Pure Appl. Opt.* **7**, 374-381 (2005).

[13] F. J. Lawrence, L. C. Botten, K. B. Dossou, R. C. McPhedran, and C. M. de Sterke, "Photonic-crystal surface modes found from impedances," *Phy. Rev.* **A82**, 053840 (2010).

>
11

| Bandgap (band) | PC_I | | | | | PC_II | | | | | PC_III | | | | |
|---|---|---|---|---|---|---|---|---|---|---|---|---|---|---|---|
| | PC_A | | PC_A' | | PC_A'A | PC_A | | PC_B | | PC_BA | PC_C | | PC_D' | | PC_D'C |
| | Zak phase | Sign of reflection phase | Zak phase | Sign of reflection phase | Interface state | Zak phase | Sign of reflection phase | Zak phase | Sign of reflection phase | Interface state | Zak phase | Sign of reflection phase | Zak phase | Sign of reflection phase | Interface state |
| 4 | | − | | − | no | | − | | + | yes | | | | | |
| (3) | 0 | | π | | | 0 | | π | | | | | | | |
| 3 | | + | | − | yes | | + | | + | no | | | | | |
| (2) | π | | 0 | | | π | | 0 | | | π | | 0 | | |
| 2 | | + | | + | no | | + | | − | yes | | + | | − | yes |
| (1) | 0 | | π | | | 0 | | π | | | 0 | | 0 | | |
| 1 | | − | | + | yes | | − | | − | no | | − | | + | yes |
| (0) | 0 | | π | | | 0 | | 0 | | | 0 | | π | | |

Table 1 : Summary of the properties of photonic crystals with interface states. Unit cells (thickness in nm) for PC_A, PC_A', PC_B, PC_C, and PC_D' are $SiO_2$-$TiO_2$-$SiO_2$ (208.5-192-208.5), $TiO_2$-$SiO_2$-$TiO_2$ (96-417-96), $SiO_2$-$TiO_2$-$SiO_2$ (139.5-289-139.5), $SiO_2$-Si-$SiO_2$ (275-114-275), and Si-$SiO_2$-Si (114.5-276-114.5), respectively.

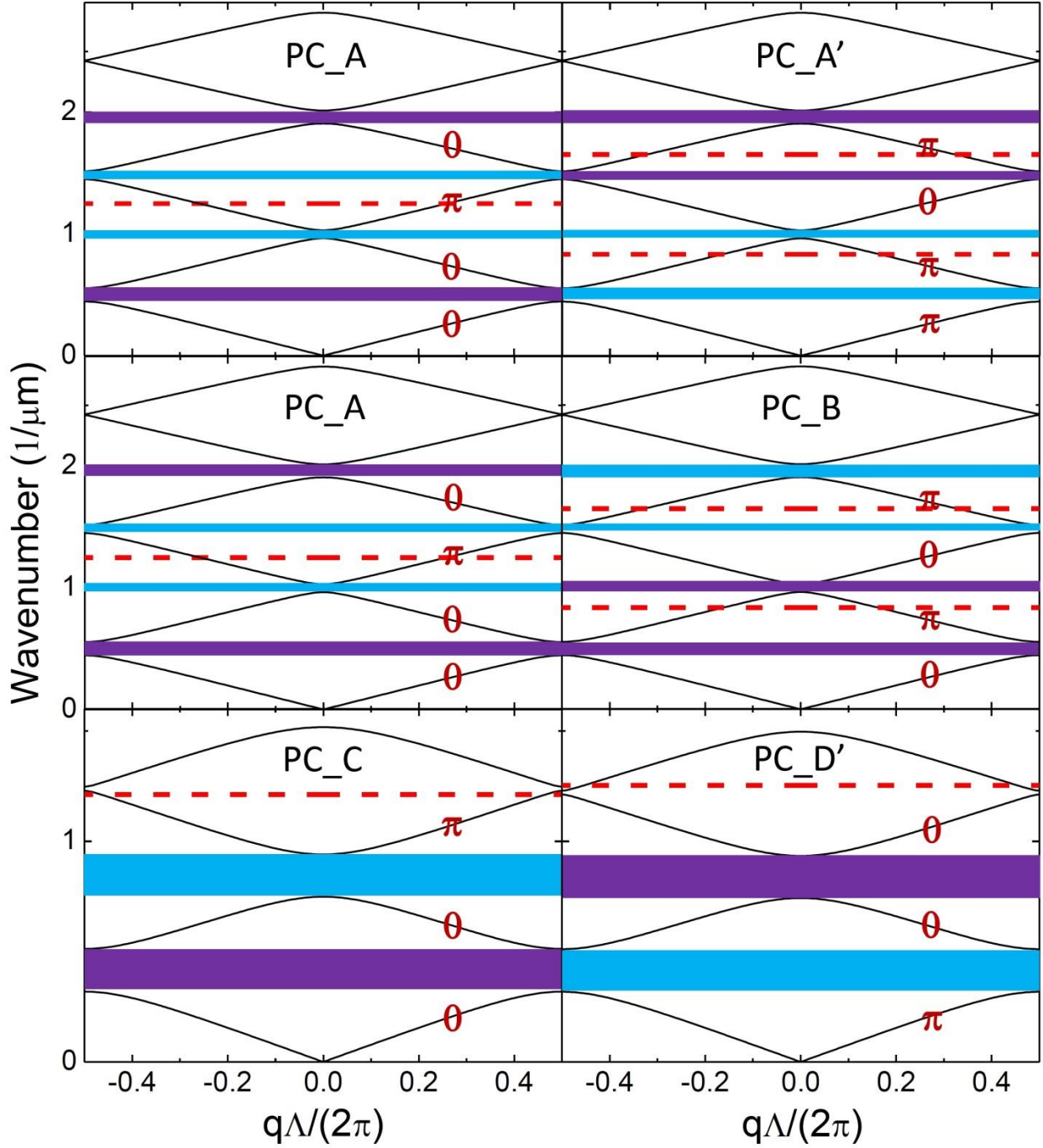

Fig. 1 : Band diagrams (wavenumber vs. $q\Lambda/2\pi$) for the PCs listed in Table 1. The Zak phase of each band below the first band crossing is labeled in red. The bandgaps below the band crossing are shown as horizontal color bars: cyan for +ve and purple for −ve reflection phase, respectively. The horizontal red dash lines correspond to wavenumbers determined by Eq. [5] for bands with Zak phase of π.

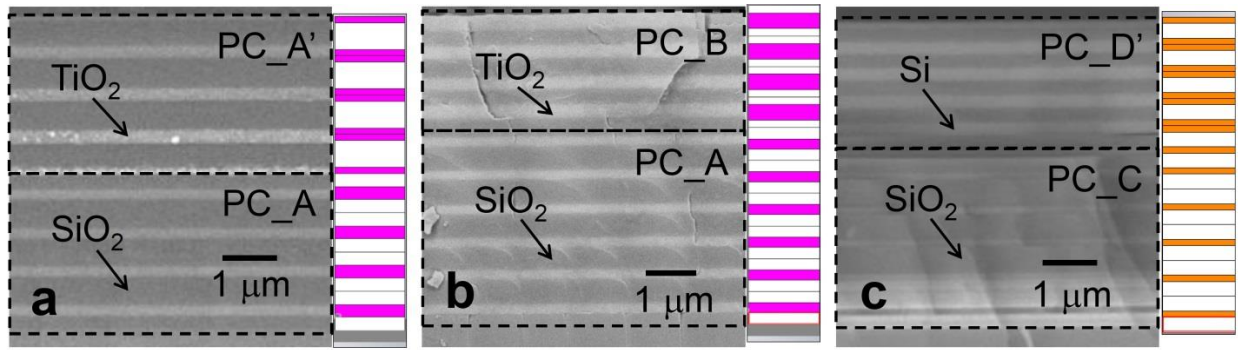

Fig. 2 : Cross section SEM images for (a) PC_I (4 unit cells PC_A' on 4 unit cells PC_A), (b) PC_II (4 unit cells PC_B on 6 unit cells PC_A), and (c) PC_III (5 unit cells PC_D' on 5 unit cells PC_C). The right schematics are unit cells of the PCs with pink color for $TiO_2$, orange color for Si, and colorless for $SiO_2$. For PC_A in (a) the average thickness is 410 nm and 211 nm for $SiO_2$ and $TiO_2$, respectively. For PC_A' in (a) the thickness is 428 nm and 173 nm for $SiO_2$ and $TiO_2$, respectively. For PC_A in (b) the average thickness is 442 nm and 195 nm for $SiO_2$ and $TiO_2$, respectively. For PC_B in (b) the thickness is 274 nm and 284 nm for $SiO_2$ and $TiO_2$, respectively. For PC_C in (c) the average thickness is 468 nm and 126 nm for $SiO_2$ and Si, respectively. For PC_D' the average thickness is 256 nm and 220 nm for $SiO_2$ and Si, respectively.

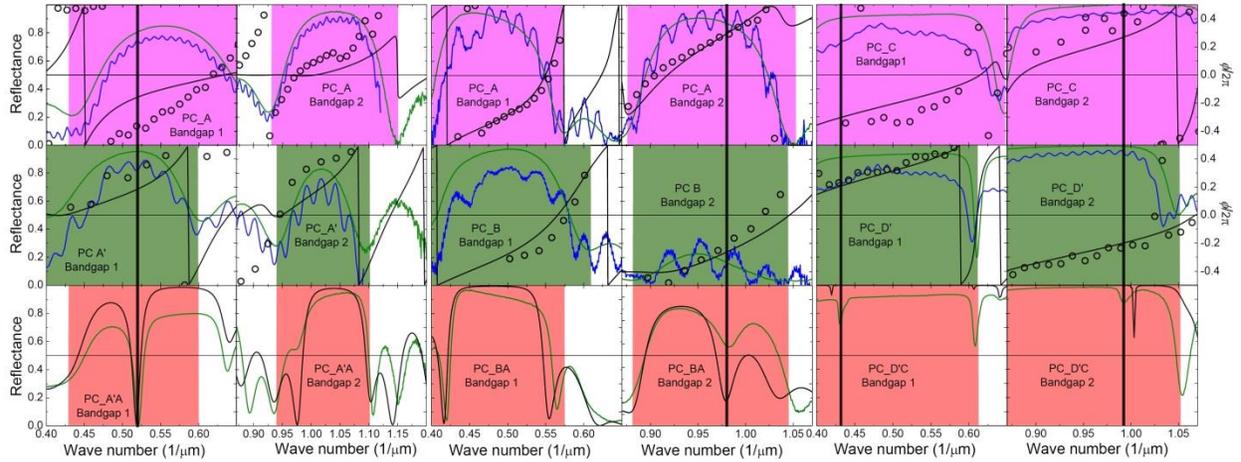

Fig. 3 : Top and middle rows are reflection spectra (blue and green curves are for samples with and without top glass, respectively) and reflection phases (black circles) for the PCs in Fig. 2 as labeled. The black curves are numerical results for reflection phases calculated using slightly adjusted thicknesses as obtained from the SEM images in Fig. 2 to account for small difference in layer thickness at the measured areas. The colored regions are bandgaps of the PCs. The last row is reflectance (in green) of the combined PCs. Odd and even columns correspond to bandgaps 1 and 2, respectively. The thick black vertical lines indicate the experimental interface states for the combined PCs. The black curves are calculated reflectance of the combined PCs.